%% file: rnaas.tex
\documentclass[twocolumn]{aastex62}
\graphicspath{{./}{figures/}}
\newcommand{\gr}{$\gamma$-ray}

\newcommand{\Fermi}{\textit{Fermi}}

\usepackage{hyperref}

\usepackage{subfiles}

\begin{document}

\title{Simultaneous Millimeter-wave, Gamma-ray, and Optical Monitoring of the Blazar PKS 2326-502 During a Flaring State}
%\input{authors}
%\author{\input{authors.tex}}

\correspondingauthor{John C.Hood II}
\email{hood.astro@gmail.com}

%\shortauthors{J.C. Hood II et al.}
%

\shorttitle{Multi-wavelength variability of PKS 2326-502}

\input{authors.tex}

\begin{abstract}
%While continuous high-cadence monitoring of active galactic nuclei (AGN) is common at \gr, optical, and radio wavelengths, AGN monitoring in the millimeter-wave (mm-wave) band has mostly been restricted to short campaigns on targeted sources. 
Including millimeter-wave (mm-wave) data in multi-wavelength studies of the variability of active galactic nuclei (AGN) can provide insights into AGN physics that are not easily accessible at other wavelengths. We demonstrate in this work the potential of cosmic microwave background (CMB) telescopes to provide long-term, high-cadence mm-wave AGN monitoring over large fractions of sky. We report on a pilot study using data from the SPTpol instrument on the South Pole Telescope (SPT), which was designed to observe the CMB at arcminute and larger angular scales. Between 2013 and 2016, SPTpol was used primarily to observe a single 500 deg$^{2}$ field, covering the entire field several times per day with detectors sensitive to radiation in bands centered at 95 and 150 GHz. We use SPT 150 GHz observations to create AGN light curves, and we compare these mm-wave light curves to those at other wavelengths, in particular \gr\ and optical. In this Letter, we focus on a single source, PKS~2326-502, which has extensive, day-timescale monitoring data in gamma-ray, optical, and now mm-wave between 2013 and 2016. We find PKS~2326-502 to be in a flaring state in the first two years of this monitoring, and we present a search for evidence of correlated variability between mm-wave, optical R band, and \gr\ observations. This pilot study is paving the way for AGN monitoring with current and upcoming CMB experiments such as SPT-3G, Simons Observatory, and CMB-S4, including multi-wavelength studies with facilities such as VRO-LSST.
\end{abstract}


\section{Introduction} \label{sec:intro}
Active galactic nuclei (AGN) are accreting supermassive ($M \gtrsim10^5 M_{\odot} $) black holes commonly found at the centers of massive galaxies \citep[e.g.,][]{kormendy95, gebhardt00}. The Unified Model of AGN proposes to explain observed categories of AGN via a scenario in which the appearance of a source depends on the angle between the axis of symmetry of the source and the line of sight of the observer \citep[e.g.,][]{antonucci93, urry95}. For example, in this scenario, blazars---radio-loud AGN\footnote{Radio-loud AGN are generally defined as AGN with a ratio of radio (5 GHz) to optical (B-band) flux $\geq$ 10 \citep{kellermann89}.} that also emit strongly in the \gr\ band—are understood to have a relativistic jet pointed at relatively small angles ($<$5 deg) to the observer.  The spectral energy distribution (SED) of blazars has a characteristic double-humped structure, with one peak located anywhere from the high-frequency radio to the soft X-ray band, caused by synchrotron emission from energetic electrons in the blazar jet, and a high-energy peak in the MeV-TeV \gr\ band \citep[e.g.,][]{fossati98}.

The source of the high-energy peak is still under debate, with models for the production of \gr\ photons classified into two broad classes: hadronic and leptonic models \citep[e.g.,][]{blandford19}. In hadronic models, processes such as photo-pion  production are responsible for the \gr\ peak, while in leptonic models, the \gr\ peak is caused by inverse-Compton scattering of lower-energy photons, which can be the same synchrotron photons responsible for the low-energy peak (the ``synchrotron self-Compton" model) or other components of the radiation field (the ``external inverse-Compton" model, e.g., \citealt{sikora94}).

A key to distinguishing between these models is what they predict for multi-wavelength observations of blazar flares. Leptonic models have been successful in explaining several observed aspects of blazars \citep{sikora94, sikora03}. The simplest interpretation of leptonic models predict that when observing AGN light curves in multiple wavelengths, there should be correlated variability between the synchrotron peak and the high-energy peak. This behavior has been observed in many cases (e.g., \citealt{bonning09}), but evidence exists that it may not always be present. For example, in the multi-wavelength study of PKS~0208-502, an ``orphan flare” was observed, in which a significant flux increase is seen in the optical/infrared bands but not in the \gr\ band \citep{chatterjee13a,chatterjee13b}.

Multi-wavelength studies of blazar flares have traditionally included \gr, X-ray, optical, infrared, and radio emission. Since millimeter-wavelength (mm-wave) radiation is a strong tracer of synchrotron emission, observations of AGN at these wavelengths should help identify the true origin of the blazar SED. Recent studies have shown that on longer timescales, mm-wave variability is better correlated with \gr\ emission than optical \citep{meyer19,zhang22}, while on shorter timescales features tend to correlate more between the optical and \gr. This points toward the possibility of synchrotron emission produced in different regions of the blazar being responsible for the mm--\gr\ correlation and the optical--\gr\ correlation. 

It has recently been recognized that cosmic microwave background (CMB) experiments have the potential to be used as AGN monitors \citep[e.g.,][]{holder19}. AGN appear as bright point sources in maps made with CMB experiments, and current CMB experiments are sufficiently sensitive to detect many AGN at high signal-to-noise ratio (S/N) in short observations. When combined with an observing strategy that results in high-cadence observations of the same patch of sky over many years, CMB data-sets are effective for AGN monitoring.

We have undertaken a pilot study of AGN variability using mm-wave data from SPTpol, the second-generation camera on the South Pole Telescope (SPT). The SPTpol survey enables the monitoring of tens of mm-bright AGN on timescales from days to years at high S/N ($>$ 10 in a 36-hour coadd). These observations provide the opportunity to include high-cadence mm-wave data in the study of the physical mechanisms behind AGN emission. 

Although our SPTpol AGN monitoring campaign includes tens of sources, we choose to focus on the blazar PKS~2326-502 for this pilot study because of its long history of observations in multiple wavelengths \citep[e.g.,][]{dutka17}. PKS~2326-502 is among the targets of monitoring by both the \Fermi\ Large Area Telescope (LAT) and Yale Small and Moderate Aperture Research Telescope System (SMARTS) Blazar Group collaborations. In particular, PKS~2326-502 has publicly available \Fermi\ (\gr) and SMARTS (optical) observations over most of the time period over which we have SPTpol data.

\section{Observations} \label{sec:observations}
In our study of PKS~2326-502, we use data from SPT, SMARTS, and \Fermi-LAT. In this section, we describe the observations and data reduction for each instrument.

\subsection{SPT} \label{subsec:SPT}
The SPT \citep{carlstrom11} is a 10-meter telescope located at the geographic South Pole and dedicated to making low-noise, high-resolution maps of the mm-wave sky, with the primary goal of mapping the temperature and polarization anisotropies in the CMB. Three separate cameras have been installed on the telescope, each used to map multiple large patches of the Southern Celestial Hemisphere. This work uses data from the second-generation camera, SPTpol. From 2013 to 2016 SPTpol was used during most of the year to survey 500\,$\mathrm{deg^2}$ of the southern extragalactic sky at arcminute resolution to mJy noise levels in bands centered at 95 and 150~GHz. The 500\,$\mathrm{deg^2}$ SPTpol survey consists of $\sim$3500 observations of a field covering $22^h$ to $2^h$ in right ascension and -$65^\circ$ to -$50^\circ$ in declination \citep{henning18}. For this study we take 150~GHz maps made from individual observations and combine them into 36-hour bundles, which provides a reasonable match with the cadence of other datasets while also providing high S/N on a sufficient number of sources.  

Once bundle maps are created, we apply a matched filter that removes the long-wavelength modes from each map, maximizing the S/N on point sources. These filtered bundles have a 1$\sigma$ error of $\sim$9\,mJy, providing us with S/N $> 10$ on 25 AGN in the 500\,$\mathrm{deg^2}$ field. We perform a series of calibration and systematic checks for each bundle. We check and correct per-bundle astrometry by comparing the positions of bright sources to those in the AT20G catalog \citep{murphy10}. We correct the calibration of each bundle by calculating the cross-spectrum of that bundle map with the \textit{Planck} 143~GHz map \citep{planck18-1} and scaling the bundle map so that the cross-spectrum agrees with that calculated for the average of all bundles. We additionally check for contamination in each bundle (such as from sidelobe pickup from the Sun in observations during the Austral summer, which would appear as bright streaks in our observations) by visually inspecting $5^\prime \times 5^\prime$ patches of sky centered on PKS~2326-502. No such contamination was detected. Once all maps have been calibrated and checked, we extract the fluxes that are used to create the mm-wave light curve of PKS~2326-502.

\subsection{SMARTS} \label{subsec:SMARTS}
This paper makes use of optical/near-infrared light curves that are available at the SMARTS website.\footnote{\url{www.astro.yale.edu/smarts/glast/home.php}} The SMARTS telescope is located in Cerro Tololo Chile, and is thus well suited to monitoring of Southern Hemisphere targets. The SMARTS blazar sample was initially (in 2008) defined to include all \Fermi-LAT-monitored blazars on the initial public release list with declination $<$ $20^\circ$. Observations were made in the B,\ V,\ R,\ J \& K bands, with an observing cadence of 1 to 3 days. Here, we use the 1-day cadence optical R band observations to match the SPT cadence as closely as possible. The full details of the data selection and analysis procedure for SMARTS data can be found in \citet{bonning12}.

\subsection{Fermi LAT} \label{subsec:LAT}
The {\Fermi}-LAT light curve for PKS~2326-502 is taken from the {\Fermi}-LAT Light Curve Repository (LCR, \citealt{abdollahi23}).\footnote{\url{https://fermi.gsfc.nasa.gov/ssc/data/access/lat/LightCurveRepository}} The LCR is a public database of multi-cadence flux-calibrated light curves for over 1500 variable sources in the 10-year {\Fermi}-LAT point source catalog (4FGL-DR2, \citealt{ballet20}). The light curves generated by the LCR span the duration of the mission and are obtained by performing an unbinned likelihood analysis over the energy range $100\,\mathrm{MeV}–100$\,GeV. The LCR analysis uses the standard {\Fermi}-LAT Science Tools (version v11r5p3) and the P8R2\_SOURCE\_V6 instrument response functions on P8R3\_SOURCE class photons. Photons are selected from a $12^{\circ}$ region of interest (ROI) centered on the location of the 4FGL-DR2 counterpart of PKS~2326-502 (4FGL J2329.3-4955). A zenith angle cut of $90^{\circ}$ is used to prevent contamination from the Earth’s limb. Included in the photon distribution model used to calculate the flux of the target source are all 4FGL-DR2 point sources within $30^{\circ}$ as well as Galactic diffuse (gll\_iem\_v07.fits) and isotropic (iso\_P8R3\_SOURCE\_V3\_v1) background models. The LCR provides light curves in cadences of 3 days, 1 week, and 1 month. For this analysis we use the minimum available time binning of 3 days.

\section{Methods}\label{sec:methods}
In this Letter, we report both qualitative and quantitative results from the analysis of multi-wavelength light curves of PKS~2326-502.  Quantitatively, we measure the local cross-correlation functions (CCFs)\footnote{Here ``local'' refers to calculating the mean and variance of both light curves over individual time lag bins rather than the entire light curve, see \citet{welsh99} for details.} of year-long light curves and calculate the significance by comparing these to uncorrelated simulations. The simulations were created by taking the power spectrum of the light curve from each data set, fitting to a model in which the light curve fluctuation power as a function of temporal frequency $P(f) = P_0 (1 + (f/f_\mathrm{knee})^{-\alpha})$, and producing 10,000 simulations of light curves from each model power spectrum. The simulated light curves are generated in Fourier space with random phase (i.e., they obey Gaussian statistics in real space). Some recent results have indicated that, at least in the \gr, blazar variability is better described by a log-normal probability distribution than a Gaussian \citep[e.g.,][]{duda21}. We have created an alternate set of simulations with log-normal statistics and do not see any significant change in our results when we use this alternate set.

We calculate the local CCF for each pair of light curves in the real data and all 10,000 simulations using the following procedure.
For a given time lag bin $\tau$, we select all data points in light curves $a$ and $b$ that satisfy:
\begin{equation}\label{eqn:tau}
t(a) - t(b) \in \tau \pm \frac{\Delta\tau}{2},
\end{equation}
where $\Delta\tau$ is the bin width, and $t(a)$ and $t(b)$ are the times for observations in each light curve. We define the local CCF as:
\begin{equation}\label{eqn:ccf}
\mathrm{CCF}(\tau) = \frac{\sum_{i=1}^{n}(a_{i}-\bar{a})(b_{i}-\bar{b})}{(n-1)s_{a}s_{b}},
\end{equation}
where $s_{a}$ is defined as:
\begin{equation}
s_{a} = \sqrt{\frac{1}{n-1} \sum_{i=1}^{n}(a_{i}-\bar{a})^{2}},
\end{equation}
$\bar{a}$ is defined as:
\begin{equation}
\bar{a} = \frac{1}{n}\sum_{i=1}^{n}a_{i}
\end{equation}
and $i$ runs over all pairs of points that satisfy Equation~\ref{eqn:tau}. The simulated CCFs were then used to find the 1$\sigma$, 2$\sigma$ and 3$\sigma$ contours for the data CCFs.

As discussed in \citet{welsh99}, measuring the full correlation function is challenging in data that is dominated by the  longest time-scale feature in the data. We also wish to remove possible dependencies on the binning timescale of any of our data sets. For these reasons, we boxcar-smooth all of the light curves with a 7-day window, and we only calcluate the full CCF on data that has been detrended using a polynomial filter. We use a fifth-order polynomial per year, which preserves features up to time scales of months.
For data that has not been detrended, we only calculate the zero-lag correlation and associated p-value. This p-value is estimated by calculating the number of simulations that have a higher zero-lag correlation than the data:
\begingroup
\small
\begin{equation} \label{zlpv}
p(\mathrm{CCF}, \tau=0) = \frac{N(\mathrm{CCF}_\mathrm{sim}, \tau=0 > \mathrm{CCF}_\mathrm{data},\tau=0)} {N_\mathrm{sims}}
\end{equation}
\endgroup
For detrended data, we calculate this zero-lag correlation and p-value, and we further plot the full CCF and look for evidence of lags between the flaring in different bands.

\section{Results}\label{sec:results}
Multi-wavelength (\gr, optical, mm-wave) light curves for four years of monitoring of PKS~2326-502 are shown in Figure~\ref{fluxes}. 
We note that the raw statistical significance of the variability in all three bands is high: The typical S/N in a single 36-hour SPT light curve point is $\sim$50 in the quiescent state and over 200 in the flaring state, and the corresponding S/N for the 3-day \Fermi\ light curve points are 1--2 and 7--10. For SMARTS-R, where we only have data in the flaring state, the typical S/N per 1-day point is $\sim$50.

Several features of these light curves which make up the primary results of this Letter are evident by-eye in Figure~\ref{fluxes}, including:
\begin{figure*}[!htb]
    \centering  \includegraphics[width=0.85\textwidth]{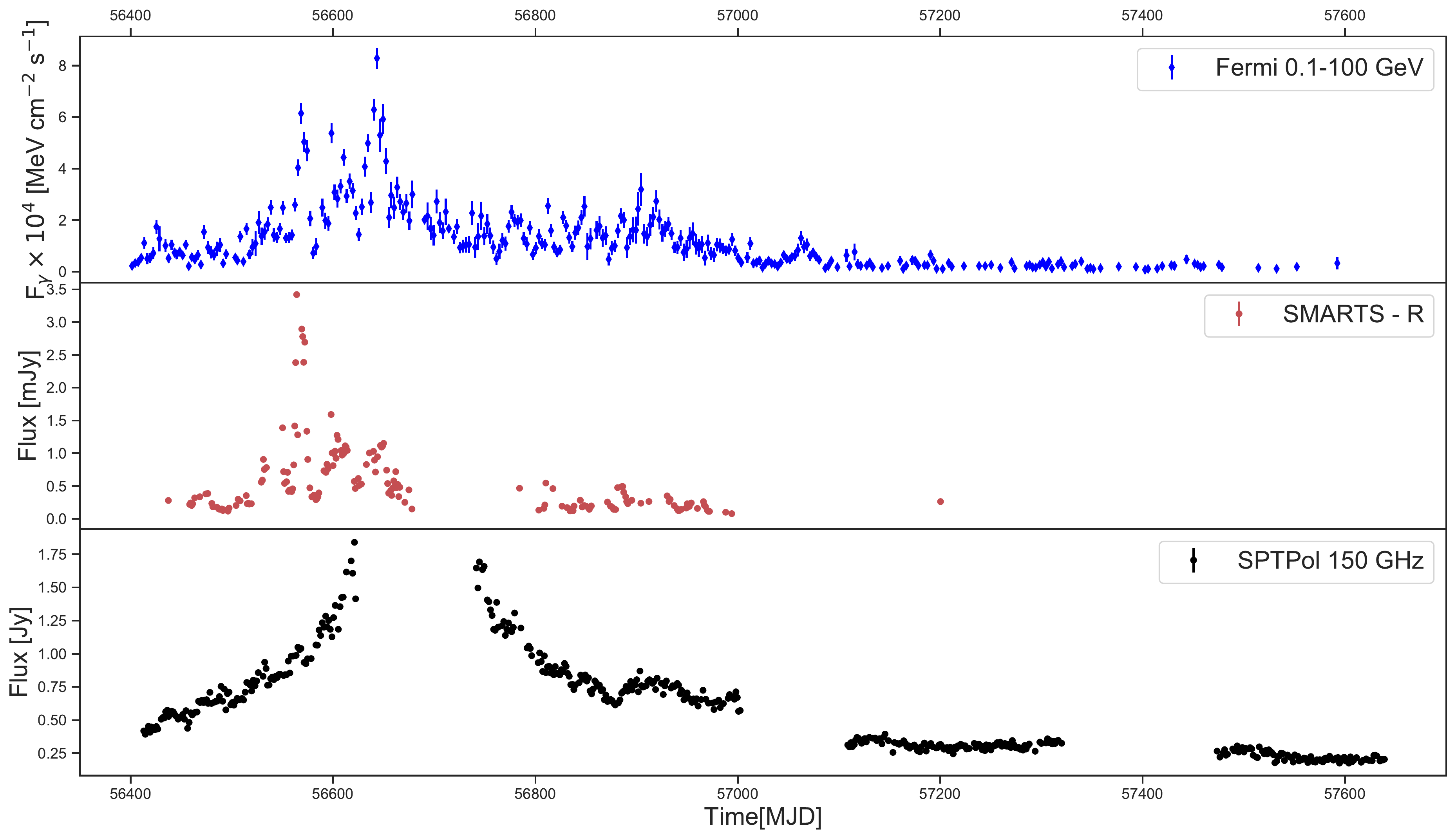}
    \caption{Light curves for PKS~2326-502. \textit{Top}: Fermi-LAT; \textit{Middle}: SMARTS optical R; \textit{Bottom}: SPTpol 150GHz. Evident by eye are long-timescale correlation between mm-wave and \gr\ observations and short-timescale correlation between optical and \gr\ observations. For reference, MJD 56400 was calendar date April, 18 2013. The time gaps in the SPTpol data are periods during the austral summer when the primary 500\,$\mathrm{deg^2}$ field was not observed to avoid solar contamination.}
    \label{fluxes}
\end{figure*}

\begin{enumerate}
\setlength{\itemsep}{0pt}
  \item A long-timescale flaring state in the first two years followed by a two-year quiescent period.
  \item Long-timescale correlation between mm-wave and \gr\ data, with the mm-wave light curve appearing to decay more slowly than the \gr\ one.
  \item Short-timescale correlation between \gr\ and optical data.
\end{enumerate}

\begin{figure*}[!htb]
    \centering
    \includegraphics[width=0.85\textwidth]{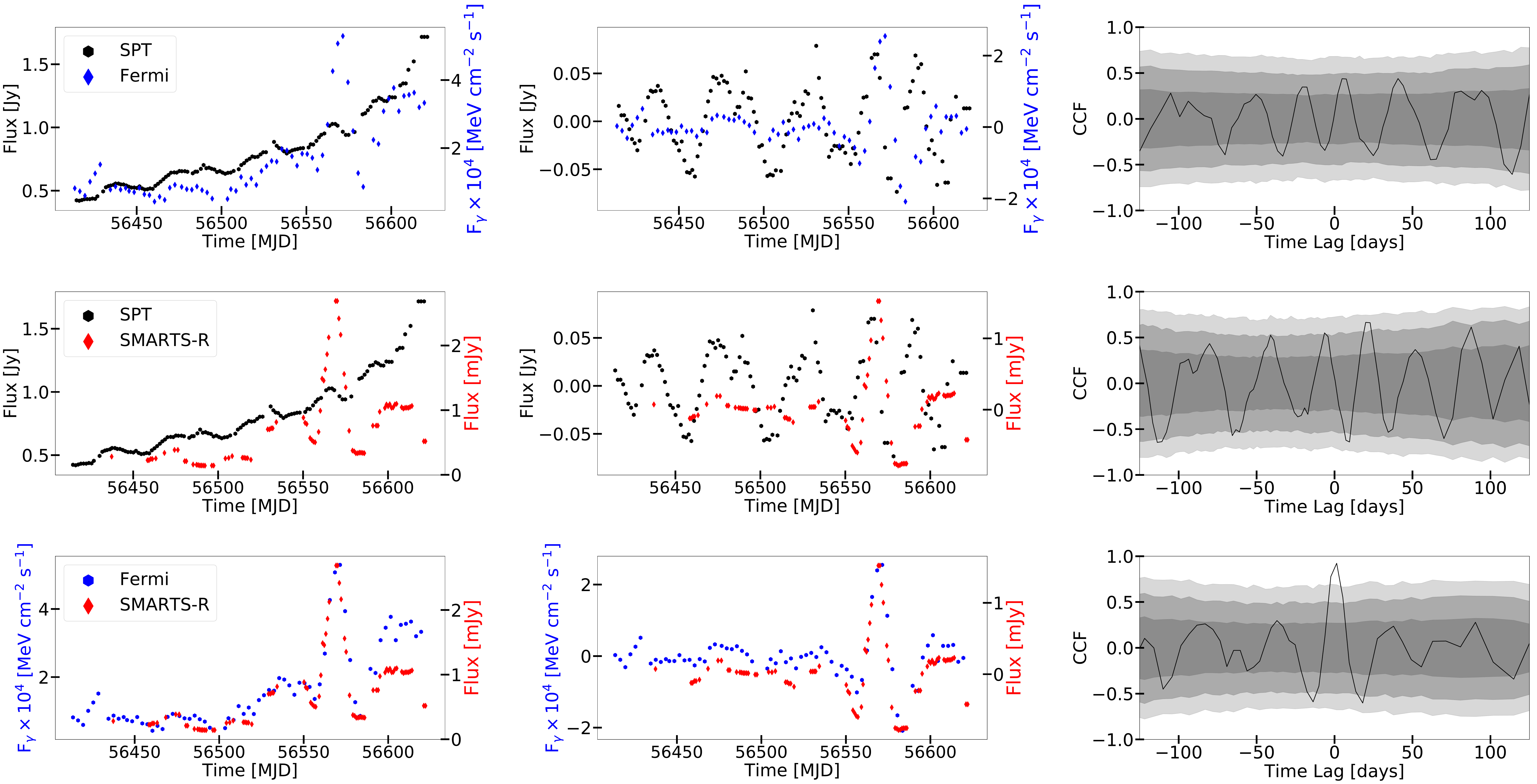}
    \caption{Year one boxcar-smoothed light curves (left column), smoothed and detrended light curves (middle column), and detrended CCFs (right column). The grey shaded regions represent the 1$\sigma$, 2$\sigma$ and 3$\sigma$ contours of simulated CCFs. }
    \label{year1}
\end{figure*}

\begin{figure*}[!htb]
    \centering
    \includegraphics[width=0.85\textwidth]{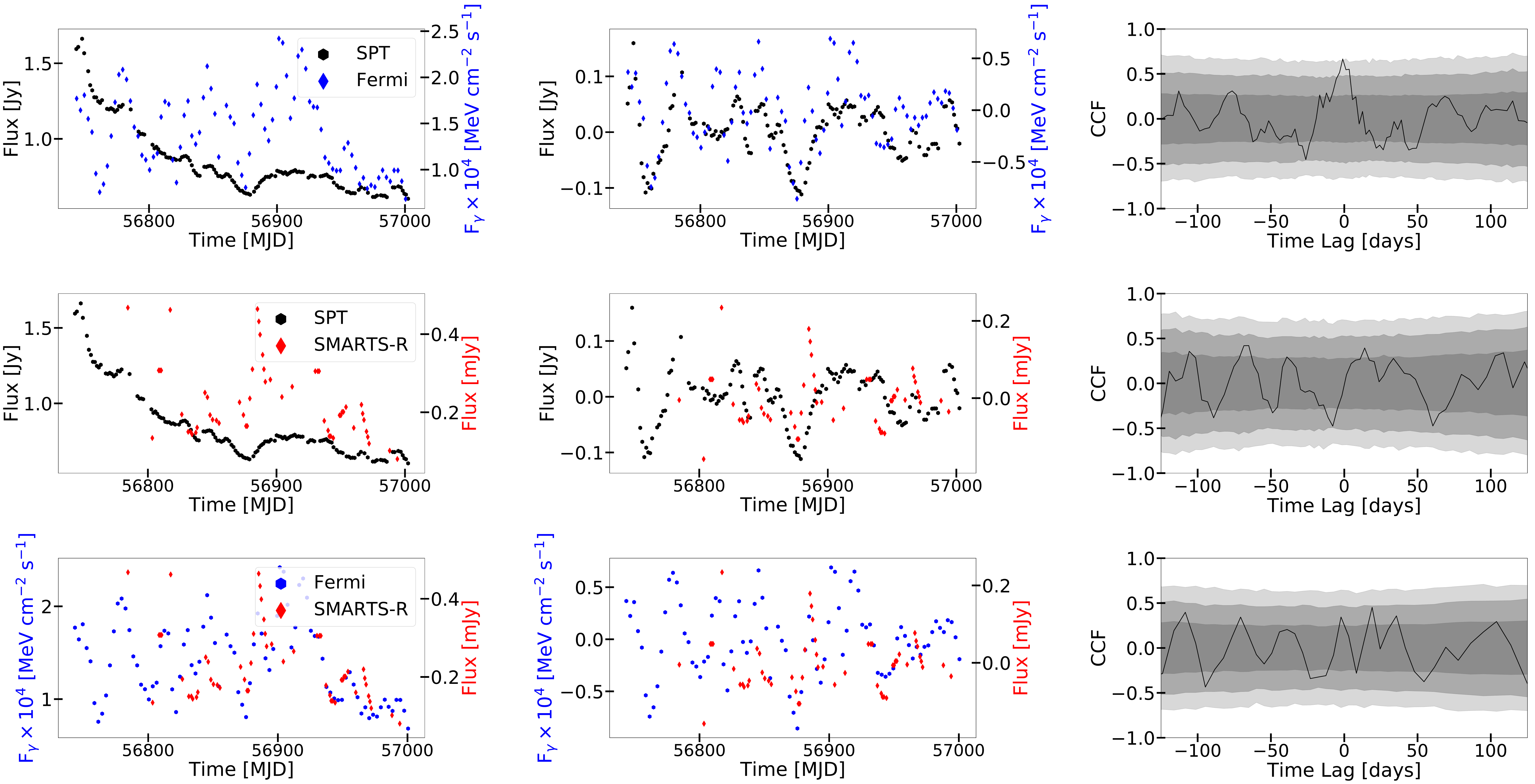}
    \caption{Same as Figure~\ref{year1} but for year 2.}
    \label{year2}
\end{figure*}

For our quantitative analyses, we focus on the observations made in the first two years of available SPTpol data (2013-2014), because: 1) PKS~2326-502 entered into a quiescent state thereafter, and 2) there is no publicly available optical data from SMARTS after 2014. 
For all possible pairs of data, two sets of light curves (boxcar-smoothed and smoothed-and-detrended) and CCFs for the smoothed and detrended data are shown for year 1 and year 2 in Figure~\ref{year1} and Figure~\ref{year2} respectively. 

As a rough measure of the significance of the correlated year time-scale flare in the \gr\ and mm-wave bands, we calculate the number of simulations that show a similar or larger flux increase over one year in those two bands. We find that only 42 out of 10,000 simulations show a factor of 2.5 increase over one year in both bands. We chose a factor of 2.5 because the ratio of the flux in the first and last month of year one was 2.7 in SPT and 3.3 in \Fermi. Therefore, we report $4.2 \times 10^{-3}$ as a raw, non-trials-corrected p-value for this long-timescale correlated flaring state. We also calculate the zero-lag correlation for the unfiltered boxcar-smoothed year one data and we find a zero-lag correlation value of 0.75 for SPT x \Fermi. Only 287 simulations show a zero-lag correlation between SPT x \Fermi\ higher than this, thus we report in Table~\ref{teststats} a p-value of 0.03 for this correlation.

Another fairly strong identifiable feature in the data is the short $\sim$week-timescale flare seen in both \Fermi\ and SMARTS but not in SPT. This leads to a significant detection of zero-lag correlation even in the non-detrended data $p(\mathrm{CCF},\tau=0) = 3 \times 10^{-4}$. Once we filter out the long time-scale features we find a zero-lag correlation value much higher than found in any of our simulations and thus report a p-value of $< 10^{-4}$ in Table~\ref{teststats}. This confirms that this correlation is being driven by the shorter-timescale feature in the \Fermi\ and SMARTS light curves shown in Figure~\ref{year1}. By contrast, when we detrend SPT x \Fermi\ data in year one, we find no evidence of correlation on shorter timescales.

In contrast to year one, for year two we find a significant correlation between the detrended SPT x \Fermi\ light curves, but none for the SMARTS x \Fermi\ light curves, as shown in Figure~\ref{year2}. We also find no significant correlation between any of the data sets in year two prior to detrending. Finally we note that we measure no significant correlation at non-zero lag for any data combination in either year.

\begin{table*}[!htb]\centering
\begin{tabular}{|l||c|c|}
\hline
\multicolumn{3}{|c|}{Light Curve Statistics}\\
\hline
Dataset  & zero-lag correlation & zero-lag p-value \\
\hline
                SPT x Fermi year one (smoothed) &                 0.75 &             $2.9\times10^{-2}$ \\
    SPT x Fermi year one (smoothed \& detrended) &                $-1.3\times10^{-2}$ &             0.52 \\
  \hline
             Smarts x Fermi year one (smoothed) &                 0.92 &             $3.0\times10^{-4}$ \\
 Smarts x Fermi year one (smoothed \& detrended) &                 0.92 &             $<10^{-4}$ \\
  \hline
               SPT x Smarts year one (smoothed) &                 0.48 &             0.16 \\
   SPT x Smarts year one (smoothed \& detrended) &                 $2.6\times10^{-2}$ &             0.47 \\
   \hline
    \hline
                SPT x Fermi year two (smoothed) &                 0.23 &             0.32 \\
    SPT x Fermi year two (smoothed \& detrended) &                 0.67 &             $1.0\times10^{-3}$ \\
     \hline
             Smarts x Fermi year two (smoothed) &                 0.54 &             $7.4\times10^{-2}$ \\
 Smarts x Fermi year two (smoothed \& detrended) &                 0.34 &             $8.7\times10^{-2}$ \\
  \hline
               SPT x Smarts year two (smoothed) &                 0.32 &             0.26 \\
   SPT x Smarts year two (smoothed \& detrended) &                $9.7\times10^{2}$&             0.36 \\
\hline   
\end{tabular}
\caption{Values of zero-lag correlation and associated p-value for smoothed and smoothed-and-detrended data. Values for years one and two are reported separately.}
\label{teststats}
\end{table*}

\section{Discussion} \label{sec:discussion}
Our study of the multi-wavelength variability of PKS~2326-502 yields four primary results:

\begin{enumerate}
\setlength{\itemsep}{0pt} 
  \item Long-timescale correlation between mm-wave and \gr\ data, with the mm-wave light curve appearing to decay more slowly than the \gr\ one.
  \item Short-timescale correlation between \gr\ and optical light curves in year one.
  \item Short-timescale correlation between \gr\ and mm-wave light curves in year two.
  \item No measurable correlation between mm-wave and optical light curves. 
\end{enumerate}

These results have implications for the production mechanism of \gr s in blazars and the structure of these systems in general. Very broadly, the correlated variability we observe between the \gr\ light curves and those in the optical and mm-wave is more consistent with leptonic models of \gr\ production than with hadronic models. While a quantitative comparison of our findings with predictions of specific leptonic models are beyond the scope of this paper, we note that simple scaling arguments predict that, in the external inverse-Compton model, the fractional amplitude of a \gr\ flare should scale linearly with the synchrotron flare amplitude. On the other hand, in the synchrotron self-Compton model, the \gr\ flare amplitude should be roughly the square of that seen in synchrotron. The long-timescale flare in year one is of similar fractional amplitude in the mm-wave and \gr\ data, lending some support to the external model. We also note the longer lifetime of the mm-wave outburst is consistent with a longer radiative lifetime of mm-wave electrons as discussed in, e.g., \citet{potter18}.

Independent of the production of \gr s, a puzzling feature of our data is the complete lack of correlation between the optical and mm-wave data. For year one, a possible explanation for this lack of correlation is that we are seeing different regions of the jet in the two bands because the mm-wave synchrotron radiation is optically thick. This would also be consistent with the long-timescale correlation between the mm-wave and \gr\ light curves and the short-timescale correlation of \gr\ and optical light curves, because we would expect to see short-timescale variability only closer to the central black hole.  In this picture, the short-timescale mm-wave and \gr\ correlation in year two is consistent with the mm-wave radiation becoming optically thin. This motivates the comparison of mm-wave optical thickness at different points in the light curve. To explore this, we extract 95~GHz SPTpol fluxes for PKS~2326-502 in a subset of observations, using procedures identical those used to extract the 150~GHz flux (Section~\ref{subsec:SPT}). We measure the mm-wave spectral index which we define through:
\begin{equation}
    S_{\nu} \propto \nu^{\alpha}; 
\end{equation}
i.e., we estimate $\alpha$ as:
\begin{equation}
    \alpha = \frac{\ln(\frac{S_{150}}{S_{95}})}{\ln(\frac{\nu_{150}}{\nu_{95}})},
\end{equation}
where $S_{95}$ and $S_{150}$ are the 95 and 150 GHz fluxes, and $\nu_{95}$ and $\nu_{150}$ are the effective band centers for a synchrotron source. We estimate $\alpha$ for four 36-hour bundles each near the three prominent features in the mm-wave light curve: the peak of the long-timescale flare in year one, the short-timescale flare in year two, and the quiescent period in year three.

We find values of the mm-wave spectral index of $\alpha = -0.24$ at the peak of the year-one flare, $\alpha = -0.52$ in the year-two flare, and $\alpha = -0.95$ in the quiescent period. These values are consistent with the picture of the mm-wave optical thickness decreasing after the year-one flare peak, thus allowing us to see farther upstream in the mm-wave in year two than in year one. What this scenario does not explain is why we do not see any correlation between the optical and mm-wave radiation in year two. It is possible that the optical synchrotron radiation tracing this activity is too faint, and that any variation in the optical flux in year two is caused by an unassociated process.

\section{Conclusion} \label{sec:conclusion}
We have presented results from a pilot study using CMB data to monitor AGN, in particular the blazar PKS~2326-502. We have correlated the mm-wave light curve from SPTpol with \gr\ data from \Fermi-LAT and optical data from SMARTS. We measured long- and short-timescale correlation between the mm-wave and \gr\ light curves, and short-timescale correlation between the optical and \gr\ light curves, but we found no measurable correlation between the mm-wave and optical light curves. These results are broadly consistent with leptonic models of \gr\ production in blazars, but they imply that the production of synchrotron emission is more complex than a single source at all wavelengths. 

While this study only used data from a single object, we have mm-wave data from many more AGN in the SPTpol survey that we will use in future investigations of multi-wavelength correlation. We will further expand this monitoring program using the yet more sensitive data from the current camera on the SPT, SPT-3G \citep{sobrin22}. 
Future experiments such as Simons Observatory \citep{ade19} and CMB-S4 \citep{abazajian19} will cover up to 70\% of the sky at nearly daily cadence with similar or even higher sensitivity. These large-footprint, high-cadence CMB surveys will be particularly well-suited for correlation with optical monitoring from VRO-LSST \citep{ivezic19}. CMB experiments are poised to become an integral part of the AGN monitoring landscape.

\acknowledgments
\section{Acknowledgements}

The South Pole Telescope program is supported by the National Science Foundation (NSF) through awards OPP-1852617 and OPP-2147371. Partial support is also provided by the Kavli Institute of Cosmological Physics at the University of Chicago.  John Hood acknowledges support from the NSF through award OPP-2219065.

\vspace{5mm}
\facilities{SPT, FERMI, SMARTS}

%\bibliography{../../BIBTEX/spt.bib}

%\begin{thebibliography}{}
%\printbibliography
%\end{thebibliography}

\end{document}

%% file: authors.tex
\shortauthors{J.~C.~Hood, A.~Simpson, et al.}
\author[0000-0003-4157-4185]{J.~C.~Hood II}\affiliation{Kavli Institute for Cosmological Physics, University of Chicago, 5640 South Ellis Avenue, Chicago, IL, USA 60637}\affiliation{Department of Astronomy and Astrophysics, University of Chicago, 5640 South Ellis Avenue, Chicago, IL, USA 60637}\affiliation{Department of Physics and Astronomy, Vanderbilt University, Nashville, TN, USA 37235}
\author{A.~Simpson}\affiliation{Department of Physics, Rensselaer Polytechnic Institute, Troy, NY 12180}
\author[0000-0002-8436-1254]{A.~McDaniel}\affiliation{Department of Physics and Astronomy, Clemson University, Clemson, SC, 29631}
\author[0000-0002-7145-1824]{A.~Foster}\affiliation{Department of Physics, Case Western Reserve University, Cleveland, OH, 44106, USA}
\author{P.~A.~R.~Ade} \affiliation{Cardiff University, Cardiff CF10 3XQ, United Kingdom}
\author{M.~Ajello}\affiliation{Department of Physics and Astronomy, Clemson University, Clemson, SC, 29631}
\author{A.~J.~Anderson} \affiliation{Fermi National Accelerator Laboratory, MS209, P.O. Box 500, Batavia, IL 60510}
\author{J.~E.~Austermann} \affiliation{NIST Quantum Devices Group, 325 Broadway Mailcode 817.03, Boulder, CO, USA 80305} \affiliation{Department of Physics, University of Colorado, Boulder, CO, USA 80309}
\author{J.~A.~Beall} \affiliation{NIST Quantum Devices Group, 325 Broadway Mailcode 817.03, Boulder, CO, USA 80305}
\author{A.~N.~Bender} \affiliation{High Energy Physics Division, Argonne National Laboratory, 9700 S. Cass Avenue, Argonne, IL, USA 60439} \affiliation{Kavli Institute for Cosmological Physics, University of Chicago, 5640 South Ellis Avenue, Chicago, IL, USA 60637}
\author[0000-0002-5108-6823]{B.~A.~Benson} \affiliation{Fermi National Accelerator Laboratory, MS209, P.O. Box 500, Batavia, IL 60510} \affiliation{Kavli Institute for Cosmological Physics, University of Chicago, 5640 South Ellis Avenue, Chicago, IL, USA 60637} \affiliation{Department of Astronomy and Astrophysics, University of Chicago, 5640 South Ellis Avenue, Chicago, IL, USA 60637}
\author[0000-0003-4847-3483]{F.~Bianchini} \affiliation{School of Physics, University of Melbourne, Parkville, VIC 3010, Australia}
\author[0000-0001-7665-5079]{L.~E.~Bleem} \affiliation{High Energy Physics Division, Argonne National Laboratory, 9700 S. Cass Avenue, Argonne, IL, USA 60439} \affiliation{Kavli Institute for Cosmological Physics, University of Chicago, 5640 South Ellis Avenue, Chicago, IL, USA 60637}
\author[0000-0002-2044-7665]{J.~E.~Carlstrom} \affiliation{Kavli Institute for Cosmological Physics, University of Chicago, 5640 South Ellis Avenue, Chicago, IL, USA 60637} \affiliation{Department of Physics, University of Chicago, 5640 South Ellis Avenue, Chicago, IL, USA 60637} \affiliation{High Energy Physics Division, Argonne National Laboratory, 9700 S. Cass Avenue, Argonne, IL, USA 60439} \affiliation{Department of Astronomy and Astrophysics, University of Chicago, 5640 South Ellis Avenue, Chicago, IL, USA 60637} \affiliation{Enrico Fermi Institute, University of Chicago, 5640 South Ellis Avenue, Chicago, IL, USA 60637}
\author{C.~L.~Chang} \affiliation{Kavli Institute for Cosmological Physics, University of Chicago, 5640 South Ellis Avenue, Chicago, IL, USA 60637} \affiliation{High Energy Physics Division, Argonne National Laboratory, 9700 S. Cass Avenue, Argonne, IL, USA 60439} \affiliation{Department of Astronomy and Astrophysics, University of Chicago, 5640 South Ellis Avenue, Chicago, IL, USA 60637}
\author{P.~Chaubal} \affiliation{School of Physics, University of Melbourne, Parkville, VIC 3010, Australia}
\author{H.~C.~Chiang} \affiliation{Department of Physics, McGill University, 3600 Rue University, Montreal, Quebec H3A 2T8, Canada} \affiliation{School of Mathematics, Statistics \& Computer Science, University of KwaZulu-Natal, Durban, South Africa}
\author{T-L.~Chou} \affiliation{Kavli Institute for Cosmological Physics, University of Chicago, 5640 South Ellis Avenue, Chicago, IL, USA 60637} \affiliation{Department of Physics, University of Chicago, 5640 South Ellis Avenue, Chicago, IL, USA 60637}
\author{R.~Citron} \affiliation{University of Chicago, 5640 South Ellis Avenue, Chicago, IL, USA 60637}
\author{C.~Corbett~Moran} \affiliation{TAPIR, Walter Burke Institute for Theoretical Physics, California Institute of Technology, 1200 E California Blvd, Pasadena, CA, USA 91125}
\author[0000-0001-9000-5013]{T.~M.~Crawford} \affiliation{Kavli Institute for Cosmological Physics, University of Chicago, 5640 South Ellis Avenue, Chicago, IL, USA 60637} \affiliation{Department of Astronomy and Astrophysics, University of Chicago, 5640 South Ellis Avenue, Chicago, IL, USA 60637}
\author{A.~T.~Crites} \affiliation{Kavli Institute for Cosmological Physics, University of Chicago, 5640 South Ellis Avenue, Chicago, IL, USA 60637} \affiliation{Department of Astronomy and Astrophysics, University of Chicago, 5640 South Ellis Avenue, Chicago, IL, USA 60637} \affiliation{Dunlap Institute for Astronomy \& Astrophysics, University of Toronto, 50 St George St, Toronto, ON, M5S 3H4, Canada} \affiliation{Department of Astronomy \& Astrophysics, University of Toronto, 50 St George St, Toronto, ON, M5S 3H4, Canada}
\author{T.~de~Haan} \affiliation{Institute of Particle and Nuclear Studies (IPNS), High Energy Accelerator Research Organization (KEK), Tsukuba, Ibaraki 305-0801, Japan} \affiliation{International Center for Quantum-field Measurement Systems for Studies of the Universe and Particles (QUP), High Energy Accelerator Research Organization (KEK), Tsukuba, Ibaraki 305-0801, Japan}
\author{M.~A.~Dobbs} \affiliation{Department of Physics, McGill University, 3600 Rue University, Montreal, Quebec H3A 2T8, Canada} \affiliation{Canadian Institute for Advanced Research, CIFAR Program in Gravity and the Extreme Universe, Toronto, ON, M5G 1Z8, Canada}
\author{W.~Everett} \affiliation{Department of Astrophysical and Planetary Sciences, University of Colorado, Boulder, CO, USA 80309}
\author{J.~Gallicchio} \affiliation{Kavli Institute for Cosmological Physics, University of Chicago, 5640 South Ellis Avenue, Chicago, IL, USA 60637} \affiliation{Harvey Mudd College, 301 Platt Blvd., Claremont, CA 91711}
\author{E.~M.~George} \affiliation{European Southern Observatory, Karl-Schwarzschild-Str. 2, 85748 Garching bei M\"{u}nchen, Germany} \affiliation{Department of Physics, University of California, Berkeley, CA, USA 94720}
\author{N.~Gupta} \affiliation{CSIRO Space \& Astronomy, PO Box 1130, Bentley WA 6102, Australia}
\author{N.~W.~Halverson} \affiliation{Department of Astrophysical and Planetary Sciences, University of Colorado, Boulder, CO, USA 80309} \affiliation{Department of Physics, University of Colorado, Boulder, CO, USA 80309}
\author{G.~C.~Hilton} \affiliation{NIST Quantum Devices Group, 325 Broadway Mailcode 817.03, Boulder, CO, USA 80305}
\author[0000-0002-0463-6394]{G.~P.~Holder} \affiliation{Astronomy Department, University of Illinois at Urbana-Champaign, 1002 W. Green Street, Urbana, IL 61801, USA} \affiliation{Department of Physics, University of Illinois Urbana-Champaign, 1110 W. Green Street, Urbana, IL 61801, USA} \affiliation{Canadian Institute for Advanced Research, CIFAR Program in Gravity and the Extreme Universe, Toronto, ON, M5G 1Z8, Canada}
\author{W.~L.~Holzapfel} \affiliation{Department of Physics, University of California, Berkeley, CA, USA 94720}
\author{J.~D.~Hrubes} \affiliation{University of Chicago, 5640 South Ellis Avenue, Chicago, IL, USA 60637}
\author{N.~Huang} \affiliation{Department of Physics, University of California, Berkeley, CA, USA 94720}
\author{J.~Hubmayr} \affiliation{NIST Quantum Devices Group, 325 Broadway Mailcode 817.03, Boulder, CO, USA 80305}
\author{K.~D.~Irwin} \affiliation{SLAC National Accelerator Laboratory, 2575 Sand Hill Road, Menlo Park, CA 94025} \affiliation{Dept. of Physics, Stanford University, 382 Via Pueblo Mall, Stanford, CA 94305}
\author{L.~Knox} \affiliation{Department of Physics, University of California, One Shields Avenue, Davis, CA, USA 95616}
\author{A.~T.~Lee} \affiliation{Department of Physics, University of California, Berkeley, CA, USA 94720} \affiliation{Physics Division, Lawrence Berkeley National Laboratory, Berkeley, CA, USA 94720}
\author{D.~Li} \affiliation{NIST Quantum Devices Group, 325 Broadway Mailcode 817.03, Boulder, CO, USA 80305} \affiliation{SLAC National Accelerator Laboratory, 2575 Sand Hill Road, Menlo Park, CA 94025}
\author{A.~Lowitz} \affiliation{Department of Astronomy and Astrophysics, University of Chicago, 5640 South Ellis Avenue, Chicago, IL, USA 60637}
\author{G.~Madejski}\affiliation{Kavli Institute for Particle Astrophysics and Cosmology, SLAC National Accelerator Laboratory, Menlo Park, CA 94025, USA}
\author{M.~Malkan}\affiliation{Department of Physics and Astronomy, University of California, Los Angeles, CA 90095-1547, USA}
\author{J.~J.~McMahon} \affiliation{Kavli Institute for Cosmological Physics, University of Chicago, 5640 South Ellis Avenue, Chicago, IL, USA 60637} \affiliation{Department of Physics, University of Chicago, 5640 South Ellis Avenue, Chicago, IL, USA 60637} \affiliation{Department of Astronomy and Astrophysics, University of Chicago, 5640 South Ellis Avenue, Chicago, IL, USA 60637}
\author{S.~S.~Meyer} \affiliation{Kavli Institute for Cosmological Physics, University of Chicago, 5640 South Ellis Avenue, Chicago, IL, USA 60637} \affiliation{Department of Physics, University of Chicago, 5640 South Ellis Avenue, Chicago, IL, USA 60637} \affiliation{Department of Astronomy and Astrophysics, University of Chicago, 5640 South Ellis Avenue, Chicago, IL, USA 60637} \affiliation{Enrico Fermi Institute, University of Chicago, 5640 South Ellis Avenue, Chicago, IL, USA 60637}
\author{J.~Montgomery} \affiliation{Department of Physics, McGill University, 3600 Rue University, Montreal, Quebec H3A 2T8, Canada}
\author{T.~Natoli} \affiliation{Department of Astronomy and Astrophysics, University of Chicago, 5640 South Ellis Avenue, Chicago, IL, USA 60637} \affiliation{Kavli Institute for Cosmological Physics, University of Chicago, 5640 South Ellis Avenue, Chicago, IL, USA 60637}
\author{J.~P.~Nibarger} \affiliation{NIST Quantum Devices Group, 325 Broadway Mailcode 817.03, Boulder, CO, USA 80305}
\author{G.~Noble} \affiliation{Department of Physics, McGill University, 3600 Rue University, Montreal, Quebec H3A 2T8, Canada}
\author{V.~Novosad} \affiliation{Materials Sciences Division, Argonne National Laboratory, 9700 S. Cass Avenue, Argonne, IL, USA 60439}
\author{Y.~Omori} \affiliation{Dept. of Physics, Stanford University, 382 Via Pueblo Mall, Stanford, CA 94305}
\author{S.~Padin} \affiliation{Kavli Institute for Cosmological Physics, University of Chicago, 5640 South Ellis Avenue, Chicago, IL, USA 60637} \affiliation{Department of Astronomy and Astrophysics, University of Chicago, 5640 South Ellis Avenue, Chicago, IL, USA 60637} \affiliation{California Institute of Technology, 1200 E. California Blvd., Pasadena CA 91125 USA}
\author{S.~Patil} \affiliation{School of Physics, University of Melbourne, Parkville, VIC 3010, Australia}
\author{C.~Pryke} \affiliation{School of Physics and Astronomy, University of Minnesota, 116 Church Street S.E. Minneapolis, MN, USA 55455}
\author[0000-0003-2226-9169]{C.~L.~Reichardt} \affiliation{School of Physics, University of Melbourne, Parkville, VIC 3010, Australia}
\author{J.~E.~Ruhl} \affiliation{Department of Physics, Case Western Reserve University, Cleveland, OH, 44106, USA} \affiliation{Center for Education and Research in Cosmology and Astrophysics, Case Western Reserve University, Cleveland, OH, USA 44106}
\author{B.~R.~Saliwanchik} \affiliation{Physics Department, Center for Education and Research in Cosmology and Astrophysics, Case Western Reserve University, Cleveland, OH, USA 44106} \affiliation{Department of Physics, Yale University, P.O. Box 208120, New Haven, CT 06520-8120}
\author{K.~K.~Schaffer} \affiliation{Kavli Institute for Cosmological Physics, University of Chicago, 5640 South Ellis Avenue, Chicago, IL, USA 60637} \affiliation{Enrico Fermi Institute, University of Chicago, 5640 South Ellis Avenue, Chicago, IL, USA 60637} \affiliation{Liberal Arts Department, School of the Art Institute of Chicago, 112 S Michigan Ave, Chicago, IL, USA 60603}
\author{C.~Sievers} \affiliation{University of Chicago, 5640 South Ellis Avenue, Chicago, IL, USA 60637}
\author{G.~Smecher} \affiliation{Department of Physics, McGill University, 3600 Rue University, Montreal, Quebec H3A 2T8, Canada} \affiliation{Three-Speed Logic, Inc., Victoria, B.C., V8S 3Z5, Canada}
\author{A.~A.~Stark} \affiliation{Harvard-Smithsonian Center for Astrophysics, 60 Garden Street, Cambridge, MA, USA 02138}
\author{C.~Tucker} \affiliation{Cardiff University, Cardiff CF10 3XQ, United Kingdom}
\author{T.~Veach} \affiliation{Space Science and Engineering Division, Southwest Research Institute, San Antonio, TX 78238}
\author{J.~D.~Vieira} \affiliation{Astronomy Department, University of Illinois at Urbana-Champaign, 1002 W. Green Street, Urbana, IL 61801, USA} \affiliation{Department of Physics, University of Illinois Urbana-Champaign, 1110 W. Green Street, Urbana, IL 61801, USA} \affiliation{Center for AstroPhysical Surveys, National Center for Supercomputing Applications, Urbana, IL, 61801, USA}
\author{G.~Wang} \affiliation{High Energy Physics Division, Argonne National Laboratory, 9700 S. Cass Avenue, Argonne, IL, USA 60439}
\author[0000-0002-3157-0407]{N.~Whitehorn} \affiliation{Department of Physics and Astronomy, Michigan State University, 567 Wilson Road, East Lansing, MI 48824}
\author[0000-0001-5411-6920]{W.~L.~K.~Wu} \affiliation{Kavli Institute for Cosmological Physics, University of Chicago, 5640 South Ellis Avenue, Chicago, IL, USA 60637} \affiliation{SLAC National Accelerator Laboratory, 2575 Sand Hill Road, Menlo Park, CA 94025}
\author{V.~Yefremenko} \affiliation{High Energy Physics Division, Argonne National Laboratory, 9700 S. Cass Avenue, Argonne, IL, USA 60439}
\author{J.~A.~Zebrowski} \affiliation{Department of Physics, University of California, Berkeley, CA, USA 94720}
\author[0000-0003-0232-0879]{L.~Zhang} \affiliation{Department of Physics, University of California, Santa Barbara, CA 93106, USA} \affiliation{Department of Physics, University of Illinois at Urbana-Champaign, 1110 West Green St., Urbana, IL 61801, USA} \affiliation{Department of Astronomy, University of Illinois at Urbana-Champaign, 1002 West Green St., Urbana, IL 61801, USA}

%% file: rnaas.bbl
\begin{thebibliography}{0}
\expandafter\ifx\csname natexlab\endcsname\relax\def\natexlab#1{#1}\fi

\end{thebibliography}


\begin{thebibliography}{30}
\expandafter\ifx\csname natexlab\endcsname\relax\def\natexlab#1{#1}\fi

\bibitem[{Abdollahi {et~al.}(2023)}]{abdollahi23}
Abdollahi, S., {et~al.} 2023, arXiv e-prints, arXiv:2301.01607

\bibitem[{{Antonucci}(1993)}]{antonucci93}
{Antonucci}, R. 1993, \araa, 31, 473

\bibitem[{{Ballet} {et~al.}(2020){Ballet}, {Burnett}, {Digel}, \&
  {Lott}}]{ballet20}
{Ballet}, J., {Burnett}, T.~H., {Digel}, S.~W., \& {Lott}, B. 2020, arXiv
  e-prints, arXiv:2005.11208

\bibitem[{{Blandford} {et~al.}(2019){Blandford}, {Meier}, \&
  {Readhead}}]{blandford19}
{Blandford}, R., {Meier}, D., \& {Readhead}, A. 2019, \araa, 57, 467

\bibitem[{{Bonning} {et~al.}(2012){Bonning}, {Urry}, {Bailyn}, {Buxton},
  {Chatterjee}, {Coppi}, {Fossati}, {Isler}, \& {Maraschi}}]{bonning12}
{Bonning}, E., {et~al.} 2012, \apj, 756, 13

\bibitem[{{Bonning} {et~al.}(2009){Bonning}, {Bailyn}, {Urry}, {Buxton},
  {Fossati}, {Maraschi}, {Coppi}, {Scalzo}, {Isler}, \& {Kaptur}}]{bonning09}
{Bonning}, E.~W., {et~al.} 2009, \apjl, 697, L81

\bibitem[{{Carlstrom} {et~al.}(2011){Carlstrom}, {Ade}, {Aird}, {Benson},
  {Bleem}, {Busetti}, {Chang}, {Chauvin}, {Cho}, {Crawford}, {Crites}, {Dobbs},
  {Halverson}, {Heimsath}, {Holzapfel}, {Hrubes}, {Joy}, {Keisler}, {Lanting},
  {Lee}, {Leitch}, {Leong}, {Lu}, {Lueker}, {Luong-van}, {McMahon}, {Mehl},
  {Meyer}, {Mohr}, {Montroy}, {Padin}, {Plagge}, {Pryke}, {Ruhl}, {Schaffer},
  {Schwan}, {Shirokoff}, {Spieler}, {Staniszewski}, {Stark}, {Tucker},
  {Vanderlinde}, {Vieira}, \& {Williamson}}]{carlstrom11}
{Carlstrom}, J.~E., {et~al.} 2011, \pasp, 123, 568

\bibitem[{{Chatterjee} {et~al.}(2013{\natexlab{a}}){Chatterjee}, {Fossati},
  {Urry}, {Bailyn}, {Maraschi}, {Buxton}, {Bonning}, {Isler}, \&
  {Coppi}}]{chatterjee13a}
{Chatterjee}, R., {et~al.} 2013{\natexlab{a}}, \apjl, 763, L11

\bibitem[{{Chatterjee} {et~al.}(2013{\natexlab{b}}){Chatterjee}, {Nalewajko},
  \& {Myers}}]{chatterjee13b}
{Chatterjee}, R., {Nalewajko}, K., \& {Myers}, A.~D. 2013{\natexlab{b}}, \apjl,
  771, L25

\bibitem[{{CMB-S4 Collaboration} {et~al.}(2019){CMB-S4 Collaboration},
  {Abazajian}, {Addison}, {Adshead}, {Ahmed}, {Allen}, {Alonso}, {Alvarez},
  {Anderson}, {Arnold}, {Baccigalupi}, \& et~al.}]{abazajian19}
{CMB-S4 Collaboration}, {et~al.} 2019, arXiv e-prints, arXiv:1907.04473

\bibitem[{{Duda} \& {Bhatta}(2021)}]{duda21}
{Duda}, J., \& {Bhatta}, G. 2021, \mnras, 508, 1446

\bibitem[{{Dutka} {et~al.}(2017){Dutka}, {Carpenter}, {Ojha}, {Finke},
  {D'Ammando}, {Kadler}, {Edwards}, {Stevens}, {Torresi}, {Grandi}, {Nesci},
  {Krau{\ss}}, {M{\"u}ller}, {Wilms}, \& {Gehrels}}]{dutka17}
{Dutka}, M.~S., {et~al.} 2017, \apj, 835, 182

\bibitem[{{Fossati} {et~al.}(1998){Fossati}, {Maraschi}, {Celotti}, {Comastri},
  \& {Ghisellini}}]{fossati98}
{Fossati}, G., {Maraschi}, L., {Celotti}, A., {Comastri}, A., \& {Ghisellini},
  G. 1998, \mnras, 299, 433

\bibitem[{{Gebhardt} {et~al.}(2000){Gebhardt}, {Bender}, {Bower}, {Dressler},
  {Faber}, {Filippenko}, {Green}, {Grillmair}, {Ho}, {Kormendy}, {Lauer},
  {Magorrian}, {Pinkney}, {Richstone}, \& {Tremaine}}]{gebhardt00}
{Gebhardt}, K., {et~al.} 2000, \apjl, 539, L13

\bibitem[{{Henning} {et~al.}(2018){Henning}, {Sayre}, {Reichardt}, {Ade},
  {Anderson}, {Austermann}, {Beall}, {Bender}, {Benson}, {Bleem}, {Carlstrom},
  {Chang}, {Chiang}, {Cho}, {Citron}, {Corbett Moran}, {Crawford}, {Crites},
  {de Haan}, {Dobbs}, {Everett}, {Gallicchio}, {George}, {Gilbert},
  {Halverson}, {Harrington}, {Hilton}, {Holder}, {Holzapfel}, {Hoover}, {Hou},
  {Hrubes}, {Huang}, {Hubmayr}, {Irwin}, {Keisler}, {Knox}, {Lee}, {Leitch},
  {Li}, {Lowitz}, {Manzotti}, {McMahon}, {Meyer}, {Mocanu}, {Montgomery},
  {Nadolski}, {Natoli}, {Nibarger}, {Novosad}, {Padin}, {Pryke}, {Ruhl},
  {Saliwanchik}, {Schaffer}, {Sievers}, {Smecher}, {Stark}, {Story}, {Tucker},
  {Vanderlinde}, {Veach}, {Vieira}, {Wang}, {Whitehorn}, {Wu}, \&
  {Yefremenko}}]{henning18}
{Henning}, J.~W., {et~al.} 2018, \apj, 852, 97

\bibitem[{{Holder} {et~al.}(2019){Holder}, {Berger}, {Bleem}, {Crawford},
  {Scott}, \& {Whitehorn}}]{holder19}
{Holder}, G., {Berger}, E., {Bleem}, L., {Crawford}, T.~M., {Scott}, D., \&
  {Whitehorn}, N. 2019, \baas, 51, 331

\bibitem[{{Ivezi{\'c}} {et~al.}(2019){Ivezi{\'c}}, {Kahn}, {Tyson},
  {et~al.}}]{ivezic19}
{Ivezi{\'c}}, {\v{Z}}., {Kahn}, S.~M., {Tyson}, J.~A., {et~al.} 2019, \apj,
  873, 111

\bibitem[{{Kellermann} {et~al.}(1989){Kellermann}, {Sramek}, {Schmidt},
  {Shaffer}, \& {Green}}]{kellermann89}
{Kellermann}, K.~I., {Sramek}, R., {Schmidt}, M., {Shaffer}, D.~B., \& {Green},
  R. 1989, \aj, 98, 1195

\bibitem[{{Kormendy} \& {Richstone}(1995)}]{kormendy95}
{Kormendy}, J., \& {Richstone}, D. 1995, \araa, 33, 581

\bibitem[{{Meyer} {et~al.}(2019){Meyer}, {Scargle}, \& {Blandford}}]{meyer19}
{Meyer}, M., {Scargle}, J.~D., \& {Blandford}, R.~D. 2019, \apj, 877, 39

\bibitem[{{Murphy} {et~al.}(2010){Murphy}, {Sadler}, {Ekers}, {Massardi},
  {Hancock}, {Mahony}, {Ricci}, {Burke-Spolaor}, {Calabretta}, {Chhetri}, {de
  Zotti}, {Edwards}, {Ekers}, {Jackson}, {Kesteven}, {Lindley}, {Newton-McGee},
  {Phillips}, {Roberts}, {Sault}, {Staveley-Smith}, {Subrahmanyan}, {Walker},
  \& {Wilson}}]{murphy10}
{Murphy}, T., {et~al.} 2010, \mnras, 402, 2403

\bibitem[{{Planck Collaboration} {et~al.}(2020){Planck Collaboration},
  {Aghanim}, {Akrami}, {Arroja}, {Ashdown}, {Aumont}, {Baccigalupi},
  {Ballardini}, {Banday}, {Barreiro}, {Bartolo}, {Basak}, {Battye}, {Benabed},
  {Bernard}, {Bersanelli}, {Bielewicz}, {Bock}, {Bond}, {Borrill}, {Bouchet},
  {Boulanger}, {Bucher}, {Burigana}, {Butler}, {Calabrese}, {Cardoso},
  {Carron}, {Casaponsa}, {Challinor}, {Chiang}, {Colombo}, {Combet},
  {Contreras}, {Crill}, {Cuttaia}, {de Bernardis}, {de Zotti}, {Delabrouille},
  {Delouis}, {D{\'e}sert}, {Di Valentino}, {Dickinson}, {Diego}, {Donzelli},
  {Dor{\'e}}, {Douspis}, {Ducout}, {Dupac}, {Efstathiou}, {Elsner},
  {En{\ss}lin}, {Eriksen}, {Falgarone}, {Fantaye}, {Fergusson},
  {Fernandez-Cobos}, {Finelli}, {Forastieri}, {Frailis}, {Franceschi},
  {Frolov}, {Galeotta}, {Galli}, {Ganga}, {G{\'e}nova-Santos}, {Gerbino},
  {Ghosh}, {Gonz{\'a}lez-Nuevo}, {G{\'o}rski}, {Gratton}, {Gruppuso},
  {Gudmundsson}, {Hamann}, {Hand ley}, {Hansen}, {Helou}, {Herranz},
  {Hildebrandt}, {Hivon}, {Huang}, {Jaffe}, {Jones}, {Karakci}, {Keih{\"a}nen},
  {Keskitalo}, {Kiiveri}, {Kim}, {Kisner}, {Knox}, {Krachmalnicoff}, {Kunz},
  {Kurki-Suonio}, {Lagache}, {Lamarre}, {Langer}, {Lasenby}, {Lattanzi},
  {Lawrence}, {Le Jeune}, {Leahy}, {Lesgourgues}, {Levrier}, {Lewis},
  {Liguori}, {Lilje}, {Lilley}, {Lindholm}, {L{\'o}pez-Caniego}, {Lubin}, {Ma},
  {Mac{\'\i}as-P{\'e}rez}, {Maggio}, {Maino}, {Mand olesi}, {Mangilli},
  {Marcos-Caballero}, {Maris}, {Martin}, {Martinelli},
  {Mart{\'\i}nez-Gonz{\'a}lez}, {Matarrese}, {Mauri}, {McEwen}, {Meerburg},
  {Meinhold}, {Melchiorri}, {Mennella}, {Migliaccio}, {Millea}, {Mitra},
  {Miville-Desch{\^e}nes}, {Molinari}, {Moneti}, {Montier}, {Morgante}, {Moss},
  {Mottet}, {M{\"u}nchmeyer}, {Natoli}, {N{\o}rgaard-Nielsen}, {Oxborrow},
  {Pagano}, {Paoletti}, {Partridge}, {Patanchon}, {Pearson}, {Peel}, {Peiris},
  {Perrotta}, {Pettorino}, {Piacentini}, {Polastri}, {Polenta}, {Puget},
  {Rachen}, {Reinecke}, {Remazeilles}, {Renault}, {Renzi}, {Rocha}, {Rosset},
  {Roudier}, {Rubi{\~n}o-Mart{\'\i}n}, {Ruiz-Granados}, {Salvati}, {Sandri},
  {Savelainen}, {Scott}, {Shellard}, {Shiraishi}, {Sirignano}, {Sirri},
  {Spencer}, {Sunyaev}, {Suur-Uski}, {Tauber}, {Tavagnacco}, {Tenti},
  {Terenzi}, {Toffolatti}, {Tomasi}, {Trombetti}, {Valiviita}, {Van Tent},
  {Vibert}, {Vielva}, {Villa}, {Vittorio}, {Wand elt}, {Wehus}, {White},
  {White}, {Zacchei}, \& {Zonca}}]{planck18-1}
{Planck Collaboration}, {et~al.} 2020, \aap, 641, A1

\bibitem[{{Potter}(2018)}]{potter18}
{Potter}, W.~J. 2018, \mnras, 473, 4107

\bibitem[{{Sikora} {et~al.}(1994){Sikora}, {Begelman}, \& {Rees}}]{sikora94}
{Sikora}, M., {Begelman}, M.~C., \& {Rees}, M.~J. 1994, \apj, 421, 153

\bibitem[{{Sikora} \& {Madejski}(2003)}]{sikora03}
{Sikora}, M., \& {Madejski}, G.~M. 2003, in Astronomical Society of the Pacific
  Conference Series, Vol. 290, Active Galactic Nuclei: From Central Engine to
  Host Galaxy, ed. S.~{Collin}, F.~{Combes}, \& I.~{Shlosman}, 287

\bibitem[{{Simons Observatory Collaboration} {et~al.}(2019){Simons Observatory
  Collaboration}, {Ade}, {Aguirre}, {Ahmed}, {Aiola}, {Ali}, {Alonso},
  {Alvarez}, {Arnold}, {Ashton}, {Austermann}, {Awan}, {Baccigalupi},
  {Baildon}, {Barron}, {Battaglia}, {Battye}, {Baxter}, {Bazarko}, {Beall},
  {Bean}, {Beck}, {Beckman}, {Beringue}, {Bianchini}, {Boada}, {Boettger},
  {Bond}, {Borrill}, {Brown}, {Bruno}, {Bryan}, {Calabrese}, {Calafut},
  {Calisse}, {Carron}, {Challinor}, {Chesmore}, {Chinone}, {Chluba}, {Cho},
  {Choi}, {Coppi}, {Cothard}, {Coughlin}, {Crichton}, {Crowley}, {Crowley},
  {Cukierman}, {D'Ewart}, {D{\"u}nner}, {de Haan}, {Devlin}, {Dicker},
  {Didier}, {Dobbs}, {Dober}, {Duell}, {Duff}, {Duivenvoorden}, {Dunkley},
  {Dusatko}, {Errard}, {Fabbian}, {Feeney}, {Ferraro}, {Flux{\`a}}, {Freese},
  {Frisch}, {Frolov}, {Fuller}, {Fuzia}, {Galitzki}, {Gallardo}, {Tomas Galvez
  Ghersi}, {Gao}, {Gawiser}, {Gerbino}, {Gluscevic}, {Goeckner-Wald}, {Golec},
  {Gordon}, {Gralla}, {Green}, {Grigorian}, {Groh}, {Groppi}, {Guan},
  {Gudmundsson}, {Han}, {Hargrave}, {Hasegawa}, {Hasselfield}, {Hattori},
  {Haynes}, {Hazumi}, {He}, {Healy}, {Henderson}, {Hervias-Caimapo}, {Hill},
  {Hill}, {Hilton}, {Hilton}, {Hincks}, {Hinshaw}, {Hlo{\v{z}}ek}, {Ho}, {Ho},
  {Howe}, {Huang}, {Hubmayr}, {Huffenberger}, {Hughes}, {Ijjas}, {Ikape},
  {Irwin}, {Jaffe}, {Jain}, {Jeong}, {Kaneko}, {Karpel}, {Katayama}, {Keating},
  {Kernasovskiy}, {Keskitalo}, {Kisner}, {Kiuchi}, {Klein}, {Knowles},
  {Koopman}, {Kosowsky}, {Krachmalnicoff}, {Kuenstner}, {Kuo}, {Kusaka},
  {Lashner}, {Lee}, {Lee}, {Leon}, {Leung}, {Lewis}, {Li}, {Li}, {Limon},
  {Linder}, {Lopez-Caraballo}, {Louis}, {Lowry}, {Lungu}, {Madhavacheril},
  {Mak}, {Maldonado}, {Mani}, {Mates}, {Matsuda}, {Maurin}, {Mauskopf}, {May},
  {McCallum}, {McKenney}, {McMahon}, {Meerburg}, {Meyers}, {Miller},
  {Mirmelstein}, {Moodley}, {Munchmeyer}, {Munson}, {Naess}, {Nati},
  {Navaroli}, {Newburgh}, {Nguyen}, {Niemack}, {Nishino}, {Orlowski-Scherer},
  {Page}, {Partridge}, {Peloton}, {Perrotta}, {Piccirillo}, {Pisano},
  {Poletti}, {Puddu}, {Puglisi}, {Raum}, {Reichardt}, { }, {Rephaeli},
  {Riechers}, {Rojas}, {Roy}, {Sadeh}, {Sakurai}, {Salatino}, {Sathyanarayana
  Rao}, {Schaan}, {Schmittfull}, {Sehgal}, {Seibert}, {Seljak}, {Sherwin},
  {Shimon}, {Sierra}, {Sievers}, {Sikhosana}, {Silva-Feaver}, {Simon},
  {Sinclair}, {Siritanasak}, {Smith}, {Smith}, {Spergel}, {Staggs}, {Stein},
  {Stevens}, {Stompor}, {Suzuki}, {Tajima}, {Takakura}, {Teply}, {Thomas},
  {Thorne}, {Thornton}, {Trac}, {Tsai}, {Tucker}, {Ullom}, {Vagnozzi}, {van
  Engelen}, {Van Lanen}, {Van Winkle}, {Vavagiakis}, {Verg{\`e}s}, {Vissers},
  {Wagoner}, {Walker}, {Ward}, {Westbrook}, {Whitehorn}, {Williams},
  {Williams}, {Wollack}, {Xu}, {Yu}, {Yu}, {Zago}, {Zhang}, {Zhu}, \& {The
  Simons Observatory collaboration}}]{ade19}
{Simons Observatory Collaboration}, {et~al.} 2019, \jcap, 2019, 056

\bibitem[{{Sobrin} {et~al.}(2022){Sobrin}, {Anderson}, {Bender}, {Benson},
  {Dutcher}, {Foster}, {Goeckner-Wald}, {Montgomery}, {Nadolski}, {Rahlin},
  {Ade}, {Ahmed}, {Anderes}, {Archipley}, {Austermann}, {Avva}, {Aylor},
  {Balkenhol}, {Barry}, {Thakur}, {Benabed}, {Bianchini}, {Bleem}, {Bouchet},
  {Bryant}, {Byrum}, {Carlstrom}, {Carter}, {Cecil}, {Chang}, {Chaubal},
  {Chen}, {Cho}, {Chou}, {Cliche}, {Crawford}, {Cukierman}, {Daley}, {Haan},
  {Denison}, {Dibert}, {Ding}, {Dobbs}, {Everett}, {Feng}, {Ferguson}, {Fu},
  {Galli}, {Gambrel}, {Gardner}, {Gualtieri}, {Guns}, {Gupta}, {Guyser},
  {Halverson}, {Harke-Hosemann}, {Harrington}, {Henning}, {Hilton}, {Hivon},
  {Holder}, {Holzapfel}, {Hood}, {Howe}, {Huang}, {Irwin}, {Jeong}, {Jonas},
  {Jones}, {Khaire}, {Knox}, {Kofman}, {Korman}, {Kubik}, {Kuhlmann}, {Kuo},
  {Lee}, {Leitch}, {Lowitz}, {Lu}, {Meyer}, {Michalik}, {Millea}, {Natoli},
  {Nguyen}, {Noble}, {Novosad}, {Omori}, {Padin}, {Pan}, {Paschos}, {Pearson},
  {Posada}, {Prabhu}, {Quan}, {Reichardt}, {Riebel}, {Riedel}, {Rouble},
  {Ruhl}, {Saliwanchik}, {Sayre}, {Schiappucci}, {Shirokoff}, {Smecher},
  {Stark}, {Stephen}, {Story}, {Suzuki}, {Tandoi}, {Thompson}, {Thorne},
  {Tucker}, {Umilta}, {Vale}, {Vanderlinde}, {Vieira}, {Wang}, {Whitehorn},
  {Wu}, {Yefremenko}, {Yoon}, \& {Young}}]{sobrin22}
{Sobrin}, J.~A., {et~al.} 2022, \apjs, 258, 42

\bibitem[{{Urry} \& {Padovani}(1995)}]{urry95}
{Urry}, C.~M., \& {Padovani}, P. 1995, \pasp, 107, 803

\bibitem[{{Welsh}(1999)}]{welsh99}
{Welsh}, W.~F. 1999, \pasp, 111, 1347

\bibitem[{{Zhang} {et~al.}(2022){Zhang}, {Vieira}, {Ajello}, {Malkan},
  {Archipley}, {Capota}, {Foster}, \& {Madejski}}]{zhang22}
{Zhang}, L., {Vieira}, J.~D., {Ajello}, M., {Malkan}, M.~A., {Archipley},
  M.~A., {Capota}, J., {Foster}, A., \& {Madejski}, G. 2022, \apj, 939, 117

\end{thebibliography}
